\documentclass[compsoc,conference,a4paper,10pt,times]{IEEEtran}

\usepackage{amsmath,amssymb,amsfonts}
\usepackage{cite}
\usepackage{color}
\usepackage[inline]{enumitem}
\usepackage{graphicx}
\usepackage[breaklinks=true]{hyperref}
\usepackage{microtype}
\usepackage{textcomp}
\usepackage{xcolor}
\usepackage{xspace}
\usepackage[T1]{fontenc}
\usepackage{algorithm}
\usepackage{algpseudocode}
\usepackage{algorithmicx}

\newcommand{\Parabreak}{1.2ex}
\newcommand{\Paragraph}[1]{\vspace{\Parabreak}\noindent\textbf{#1}}

\newif\ifcommentary
\commentarytrue

\ifcommentary
\newcommand{\todo}[1]{{\color{red}\textbf{TODO:} #1}\xspace}
\newcommand{\fixme}[1]{{\color{red}\textbf{!! FIXME: #1 !!}}}
\newcommand{\msm}[1]{[{\color{magenta}msm: #1}]}

\else
\newcommand{\todo}[1]{}
\newcommand{\fixme}[1]{}
\newcommand{\msm}[1]{}
\fi

\newcommand{\atlas}{\textsl{Atlas}\xspace}

\algrenewcommand\algorithmicindent{1.0em}  

\begin{document}

\title{Atlas: A Framework for ML Lifecycle Provenance \& Transparency}

\author{\IEEEauthorblockN{Marcin Spoczynski}
\IEEEauthorblockA{\textit{Intel Labs}\\
Hillsboro, Oregon, USA\\
marcin.spoczynski@intel.com}
\and
\IEEEauthorblockN{Marcela S. Melara}
\IEEEauthorblockA{\textit{Intel Labs}\\
Hillsboro, Oregon, USA\\
marcela.melara@intel.com}
\and
\IEEEauthorblockN{Sebastian Szyller}
\IEEEauthorblockA{\textit{Intel Labs}\\
Helsinki, Finland\\
contact@sebszyller.com}
}

\maketitle

\begin{abstract}
    We propose~\atlas, a framework that enables fully attestable ML pipelines to address ML supply chain risks.
    \atlas leverages runtime pipeline monitoring and open specifications for data and software
    provenance to collect model artifact integrity
    and end-to-end lineage metadata.
    \atlas combines trusted hardware and transparency logs to enhance metadata integrity and enable efficient
    verification of ML pipeline
    operations, from training through deployment.
    Our prototype implementation of~\atlas integrates open-source tools
    to build an ML lifecycle transparency framework.
    \end{abstract}
\section{Introduction}\label{sec:introduction}

In recent years, machine learning (ML) models, have become increasingly popular.
The pervasive use of large language models (LLMs), in particular, and multi-stakeholder
involvement in model creation and deployment exacerbate security and privacy risks.
These considerations are emphasized by the global nature and the complexity of
large-scale ML deployments with different lifecycle stages (e.g., training dataset collection, execution of training).
Each stage is vulnerable to malicious or dishonest parties.
For example, data can be poisoned~\cite{biggio2012poisoning,carlini2024poisoning} during collection or training.
Service providers executing outsourced training can shorten or omit critical steps to reduce their cost.
Popular model hubs hosting pre-trained models are vulnerable to compromises that may result in corrupted, reduced, or malicious model distributions~\cite{jiang2022hub-risks,cohen2024jfrog}

On the other hand, recent regulations~\cite{eo14144,eu-ai} require model builders and other stakeholders to provide evidence of ML model security and trust.
They may need to prove low bias in their training data, offer easily verifiable performance claims, or demonstrate end-to-end integrity of model creation in high risk domains.

To address these challenges, integrity of the entire ML lifecycle must be recorded verifiably --
beginning with the data, through the training, and finally, the evaluation and deployment.
Was the data modified?
Did the hardware and software environment adhere to the specification?
Did the contractor follow the specified training procedure?
Can I trust the evaluation?
How can I guarantee that I am interacting with the intended model?
These are example questions that showcase the breadth of the involved challenges that must be tackled to provide end-to-end security.

We introduce \atlas, a framework for enhancing the security and transparency of the lifecycle of ML models.
\atlas establishes the baseline of fundamental components and capabilities needed for comprehensive provenance tracking
at each stage of the ML lifecycle.
Thus, rather than preventing attacks entirely, \atlas detects tampering by verifying the ML lifecycle.

\atlas addresses two challenges unique to ML lifecycle transparency.
First, in contrast to the clear dependency trees of traditional software, datasets and algorithm code are tightly coupled in ML models, creating significantly more intricate provenance graphs~\cite{jiang2022hub-risks}.
Attesting the integrity of these relationships requires mechanisms that can track cross-organizational transformations across heterogeneous artifacts of varying sizes and formats~\cite{ozga2021perun}.

Hence, \atlas monitors ML pipelines during execution and automatically collects ML system information using several data and software provenance frameworks.
To strengthen the integrity of provenance generation, \atlas relies on hardware trusted execution environments.

Second, models are not static binaries -- they can be further customized by downstream users.
That is, ML models undergo a non-linear lifecycle where deployment results often necessitate refinement, creating feedback loops between inferencing and data processing~\cite{ashmore2022assurance-survey}. 

This iterative process requires special adaptations to provenance tracking mechanisms. \atlas represents these non-linear development paths and enables cryptographic auditing using Merkle trees~\cite{merkle-tree1987}.


We claim the following contributions:
\begin{enumerate}[label=\arabic*.]\label{sec:introduction:contributions}
    \item We introduce \atlas, a framework designed for end-to-end ML lifecycle transparency.
    \item We instantiate \atlas using Intel Trust Domain eXtensions~\cite{tdx} and metadata-based provenance tracking.
    \item We evaluate our \atlas prototype through a fine-tuning case study with a BERT model~\cite{lin2023metabert, lin2023metabertimpl}.
\end{enumerate}
\section{Background \& Related Work}\label{sec:background-related}

\noindent\textbf{Data Provenance \& Authenticity.}
Provenance and attribution of media has recently received attention
due to the online proliferation of manipulated or forged
content using generative ML models~\cite{feng2023examining, england2021amp}.
Prior work relies primarily on cryptographic hashing and digital
signatures to provide data authenticity and integrity.

The Coalition for Content Provenance and Authenticity (C2PA)
specification~\cite{iso2024c2pa, c2pa2024spec}
digitally signs assertions about content origin to provide tamper-evident
data audit trails~\cite{laurie2023c2pa, sidnam-mauch2022usable}.

C2PA's extensible metadata format also makes it suitable for ML model artifacts~\cite{collomosse2024content}.
 
The Open Source Security Foundation (OpenSSF) Model Signing
project~\cite{ossf-model-signing2024} is a parallel effort 
focusing on integrity and authenticity of trained models.
Other prior work~\cite{england2021amp,news-provenance-project,lakefs2025}
builds upon hashing and signing with
distributed ledger technologies to create transparent and immutable content or provenance records.

These techniques are crucial building blocks for verifying ML model artifact
authenticity and provenance, but each alone is insufficient for end-to-end
ML supply chain transparency.
In contrast, \atlas aims to integrate such techniques
into ML systems to track model artifact provenance directly where
the transformations occur.

\noindent\textbf{Supply Chain Integrity.}
Recent cybersecurity regulations~\cite{eo14028, eu-cra} have shifted industry focus toward detecting supply chain threats via software dependency tracking with SBOMs~\cite{ntia-sbom}.
Similarly, the AIBOM framework~\cite{trail-of-bits-aibom2024, manifest-aibom2023}
focuses on ML model supply chain management.

Complementing BOM, efforts like OpenSSF Supply Chain Levels for
Software Artifacts (SLSA)~\cite{slsa2025} and SPDX Build~\cite{spdx-build2023}
collect build provenance, i.e., metadata describing how a particular artifact was produced.

This approach is also being considered for ML model fine
tuning~\cite{slsa-for-models2024}.
Building on such supply chain metadata efforts, a number of
frameworks~\cite{torres2019,sigstore2025,scitt2024} provide mechanisms for
collecting, digitally signing and verifying authenticated claims \emph{across}
supply chain steps.

\atlas borrows concepts from
supply chain integrity to support multiple types of software artifact provenance
at any stage of the ML lifecyle, providing a more comprehensive view of a
model's supply chain.
Works about evidence for other properties of the ML lifecycle such as
assurance~\cite{ashmore2022assurance-survey} are complementary.

\noindent\textbf{Model Lineage Tracking.}

The EQTY Lineage Explorer~\cite{eqty2023} tracks model
artifacts throughout the training process, capturing relationships between
datasets, model checkpoints and hyperparameters.
However, unlike \atlas, it lacks cryptographic authenticity properties and
focuses primarily on manually collected development-time lineage, rather than
automatically capturing and linking information across the entire ML lifecycle.

ML experiment trackers like Weights and Biases~\cite{wandb2023},
Neptune~\cite{neptune2025} and Kubeflow Pipelines~\cite{kubeflow-pipelines}
offer detailed run-time logging of model metadata about training runs, metrics, and model artifacts.
These tools do not integrate transparently with common ML frameworks, and they
typically provide only unauthenticated metadata.
\atlas, on the other hand, seeks to make model lineage verifiable and
support integration into ML frameworks like PyTorch~\cite{pytorch}.

\noindent\textbf{Hardware-Based Security for ML.}
Recent developments in trusted execution environment (TEEs) technologies
have made it more practical to run large-scale systems and workloads~\cite{akkus2024duet,galanou2023cvm},
including ML pipelines.

Chrapek et al.~\cite{chrapek2024fortify} deployed and optimized a large language model (LLM)
inside a TEE, showing how secure enclaves help protect LLM code and
data while \emph{in use}.
They maintain practical performance in two TEE configurations
based on Intel Software Guard eXtensions (Intel SGX)~\cite{sgx2013} and Intel
Trust Domain eXtensions (Intel TDX)~\cite{tdx}.

Laminator~\cite{duddu2024} and PraaS~\cite{akkus2024praas} demonstrate the application
of TEEs to ML model or dataset property attestation and verification.
Several efforts~\cite{ozga2021perun,azure-conf-inferencing2024,edgeless2025private-mode}
use TEEs to build confidentiality frameworks for different ML lifecycle stages.
These works are complementary to \atlas and may enable us to extend our framework.

Mo et al.'s survey~\cite{mo-sok2024} evaluates 38 works that use various TEE
implementations to enhance the privacy and integrity of ML training and
inference operations.
The survey highlights several gaps, including the protection of full ML lifecyles, which is the primary focus of \atlas.

We provide additional background in Appendix~\ref{sec:appendix:related}.

\section{System Overview \& Threat Model}\label{sec:problem}

\subsection{Terminology}\label{sec:problem:terminology}

In \atlas, an ML model is composed of several \textbf{artifacts}
that include the training dataset, ML algorithm, ML framework (e.g., PyTorch),
model configuration (e.g., hyperparameters, weights), and metadata (e.g.,
license).

The \textbf{ML lifecycle} consists of various stages, including
data preparation, training, evaluation and deployment.
A common synonym for ML lifecycle is ``ML supply chain'', so we use these
terms interchangeably in the paper.

The \textbf{ML pipeline} defines the sequence of operations or a workflow that
transforms a model artifact at a particular stage of the ML
lifecycle~\cite{google-ml-pipelines}.
To support standardization and repeatability, the
pipeline also facilitates workflow management and automation.

The \textbf{ML system} is the set of hardware and software components that
implement and execute an ML pipeline.
For example, an ML system for training may include orchestration tools, an
authentication service, storage systems, automation infrastructure, and
specialized compute hardware (e.g., GPUs, TPUs, or custom accelerators).

In \atlas, \textbf{metadata} describes two mains aspects about an ML model.
First, provenance metadata refers to the origin and history of custody of a
model artifact, including its history of transformations as it traverses the
ML lifecycle.
Pipeline metadata, in turn, describes the ML systems and specifics about
the operations that produced a model artifact.

An \textbf{attestation} in \atlas refers to any digitally signed metadata and
serves as evidence for model artifact or ML system authenticity, integrity and
provenance.
Attestations may be generated by hardware or software.

\subsection{System Model}\label{sec:problem:sys-model}

In \atlas, we target the \emph{multi-stakeholder} ML model lifecycle setting.


\subsubsection{Stakeholders}\label{sec:problem:sys-model:participants}

The \textbf{Artifact Producers} are individuals and organizations that create ML
model artifacts 
to provide or sell them to other parties.
Since the artifacts may represent intellectual property and/or make use of
personally identifying information (PII), producers have business and regulatory
reasons to preserve their confidentiality.
To save costs, artifact producers often
outsource the operation of ML systems to external service providers.

\textbf{ML-as-a-Service (MLaaS) providers} operate and maintain the compute
infrastructure needed to run ML systems.
MLaaS providers may offer ML-specific services that leverage general-purpose
compute (e.g.,~\cite{azure-mlaas,ibm-watsonx,google-vertex-ai}), or
provide special-purpose systems (e.g.,~\cite{kubeflow-pipelines, github-mlops2020})
that can build and run third-party ML systems.

A \textbf{Hub} is a system that stores and distributes model artifacts.
Thus, model pipelines typically ingest and output artifacts to and from hubs
during their execution, enabled by interfaces exposed by MLaaS providers.
Hubs may be operated by artifact producers themselves or by third-party vendors,
containing open or closed source artifacts (e.g.,~\cite{huggingface, pytorch-hub}).

A \textbf{Transparency Service} in \atlas is responsible for generating, storing
and distributing the metadata necessary to verify the authenticity, integrity
and provenance of model artifacts.
We envision model vendors and independent parties operating transparency services
in practice.

Transparency services interface with MLaaS providers through \emph{attestation
clients} that run alongside ML systems to obtain and attest provenance and pipeline metadata.
On the server side, a \emph{transparency log} contains the known good values
(i.e., golden values) of model artifacts and ML system components, submitted by
model producers and MLaaS providers, as well as attestations collected by the
clients throughout the ML lifecycle.

\textbf{Verification Services} evaluate or audit a particular ML
model's lifecycle with the goal of detecting unintended or malicious tampering
with the model at any stage.
In practice, model users, vendors or regulatory entities may operate
verification services.

Using the golden values\footnote{
	Golden values should be independently verified to establish their
	trustworthiness.
	Current approaches for auditing golden values include reproducibility~\cite{reproducible-builds}
	and endorsements~\cite{rats}. \atlas is agnostic to the chosen method and assumes that
	evidence of golden value verification can be made available via a transparency service.}
and attestations obtained from a transparency service,
a verification service evaluates each ML pipeline and artifact of interest
against a set of \emph{expectations}.
For example, a model producer may check that the MLaaS provider ran the expected
pipeline code, or a model user may verify that a fine-tuned model was produced
from the expected foundation model.

A \textbf{Model User} interacts with a model in an inferencing
pipeline, or in a \emph{downstream} ML pipeline as a dependency, such as a
fine-tuning or evaluation pipeline (see~\S\ref{sec:problem:sys-model:lifecycle}).

\subsubsection{ML Lifecycle}\label{sec:problem:sys-model:lifecycle}

In \atlas, we consider four high-level stages in the ML lifecycle.
Each builds upon the outputs and feedback from the others,
forming a continuous cycle in which models evolve based on
real-world usage.

\noindent\textbf{1. Data processing}: Raw data is collected, sanitized and
processed into smaller units (e.g., tokens) and collated into a structure
ingestable during training or evaluation.

\noindent\textbf{2. Training}: A training algorithm processes a given dataset using an ML system.
The output is an ML model.

\noindent\textbf{3. Evaluation}: Following training, 
model properties like its performance and accuracy undergo further fine-tuning and evaluation using a testing dataset. 

\noindent\textbf{4. Deployment}: After training and evaluation, an ML model
is deployed to a production system configured for inferencing.
New data obtained from clients during inference are sent back to a data processing
pipeline to enhance the training dataset and the model.
Model use must comply with local laws or corporate policies.

\subsection{Threat Model}\label{sec:problem:threat-model}

We consider an adversary whose goal is to produce a tampered artifact,
e.g., containing a hidden malicious component, so that a transparency
service generates a legitimate signature on the artifact or its metadata.

Thus, \atlas aims to \emph{detect} such tampering introduced via the ML
supply chain.\footnote{
	Analogously to software correctness (which also applies to ML algorithms and systems),
	establishing dataset benignity, model quality and safety is a
	complementary area of research that relies on certifications, e.g.,
	adversarial robustness~\cite{cohen2019randomized-smoothing}, differential
	privacy~\cite{wicker2024certified-dp}, or poisoning~\cite{steinhardt2017certified-poisoning}.
	We leave extending \atlas with such mechanisms as future work.}
Specifically, we consider tampering by MLaaS providers, hubs and artifact
producers, while model users, transparency and verification services are
trusted in \atlas.

Compromised MLaaS providers and hubs may involve malicious insiders,
or external adversaries seeking to subvert these systems
by exploiting vulnerable components.
Given their central position in the lifecycle, MLaaS providers and hubs may thus be able to
compromise model \emph{integrity} at various stages.

For example, a malicious MLaaS provider can poison the training data during the
curation step of the data processing stage leading to
backdoors.
A compromised hub may, for instance, present a dataset or model with a mismatched
signature (e.g., to a different model, or any of its component artifacts) to a
model user or MLaaS provider, so that pipelines in subsequent stages of the ML
lifecycle may ingest compromised dependencies.

As a result, these compromises propagate through the ML lifecycle if they go
undetected, ultimately leading to vulnerable ML models at the deployment stage.
This risk is exacerbated if a hub colludes with an MLaaS provider to introduce
or accept compromised ML pipeline inputs.

Artifact producers, on the other hand, may seek to compromise ML models
to bypass regulatory requirements, introduce exploitable vulnerabilities or steal
private information for profit (e.g,~\cite{xz-utils2024}).
Thus, producers may collude with other untrusted stakeholders, or intentionally
inaccurately declare their dependencies, to undermine the integrity of their artifacts.

\subsubsection{Trusted Parties}

\atlas considers the model users, transparency and verification services
in an ML lifecycle to be trusted.
We make the following assumptions about the systems
supporting these stakeholders:
\begin{enumerate*}[label=\arabic*)]
    \item the hardware and cryptographic primitives are implemented correctly
        and do not contain known vulnerabilities;
    \item a separate PKI system exists and organizations representing the stakeholders
        follow best practices for key management, network security and access control.
\end{enumerate*}

Further, running attestation clients in TEEs allows us to trust \atlas metadata
generation, or detect attempts of tampering by malicious MLaaS providers.
Similarly, model users and verification services can trust the integrity of
golden values and attestations stored in \atlas transparency logs (or detect
tampering) via their tamper-evident construction (see~\S\ref{sec:framework:transparency:log}).
Verification services are trusted to properly evaluate attestations, including their
digital signatures, against pre-specified model user expectations.

\subsubsection{Out of Scope}

Many available TEEs provide confidentiality features, but addressing PII and
model intellectual property concerns \emph{end-to-end} requires a more
comprehensive framework (e.g., ~\cite{apple-pcc2024, ozga2021perun}).
We plan to explore confidentiality within \atlas as future work.

While a critical threat to deployed AI applications, \atlas does not address
inference time black-box attacks (e.g., evasion attacks, model extraction,
membership inference) caused by malicious users; solutions to reduce
this risk~\cite{carlini2022membership,jagielski2020high,tramer2016stealing}
are complementary.

Side-channel attacks against hardware enclaves, physical attacks on hardware
infrastructure, and network-level denial of service attacks are also beyond the
scope of \atlas.
These attacks are the subject of a large body of prior work~\cite{mo-sok2024},
and these complementary security measures could be added to
deployments of \atlas.

\subsection{Design Requirements}\label{sec:problem:requirements}

We define the following integrity and operational requirements for \atlas:

\noindent\textbf{R1: Artifact tampering is detectable.} To provide model artifact
integrity, \atlas must enable verification services to detect unexpected modifications
to model artifacts.

\noindent\textbf{R2: Every model transformation is attested.}
Because adversaries may seek to tamper with model artifacts after they are
produced by a pipeline, \atlas attestation clients must record every model
transformation in authenticated metadata as evidence for the process, including
all inputs to the transformation.

\noindent\textbf{R3: Verifiable model lineage.}
\atlas verification services must be able to detect unintended/malicious changes
by MLaaS providers to the expected stages of the lifecycle
(e.g., pipelines operating out of order, or being omitted), from initial data
processing through model deployment.

\noindent\textbf{R4: Strongly isolated ML systems.}
To detect tampering with a pipeline during its execution, \atlas
must restrict access to its ML system by malicious MLaaS providers, and contain
compromises from propagating beyond the execution environment.

\noindent\textbf{R5: Pipeline agnostic.}
To facilitate adoption, \atlas must be agnostic to any ML pipeline
that integrates it.

\noindent\textbf{R6: Efficiency.}
We seek to minimize the computational and storage overheads incurred by \atlas
to enable the implementation and deployment of \atlas in ML systems using
commodity platforms and services.

\section{Atlas Framework}\label{sec:framework}

\atlas introduces two core components to the ML lifecycle:
\begin{enumerate*}[label=\arabic*)]
    \item the transparency service interacting with MLaaS providers;
    \item the verification service for validating model integrity and provenance.
\end{enumerate*}

\begin{figure}[h]
	\centering
	\includegraphics[width=\columnwidth]{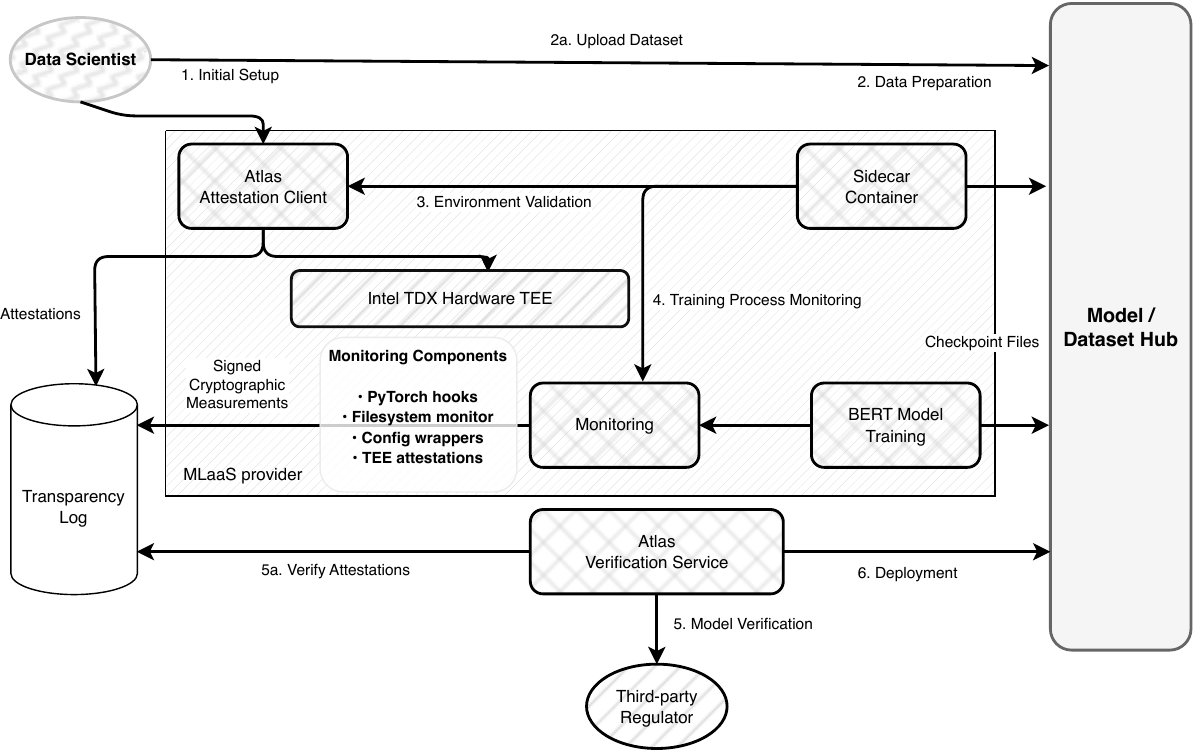}
	\caption{\atlas workflow for ML lifecycle transparency in a BERT Meta~\cite{lin2023metabert, lin2023metabertimpl} fine-tuning process. The attestation client monitors ML systems running in TEEs throughout the pipeline stages, collecting provenance metadata from initial deployment through verification.}
	\label{fig:atlas-workflow}
\end{figure}

The core techniques underlying the transparency and verification services are designed
to be general, allowing them to remain agnostic to the particular ML lifecycle
stage or pipeline they are applied to (\textbf{R5}).
Fig.~\ref{fig:atlas-workflow} depicts an example ML model lifecycle with \atlas.

\subsection{Attestation Client}\label{sec:framework:transparency:client}

\atlas combines mechanisms for artifact and runtime environment
integrity to provide transparency across the stages of the ML lifecycle.
Thus, MLaaS providers integrate an \atlas attestation client with an ML system,
each running within a dedicated trusted execution environment (TEE) (e.g., Intel
TDX~\cite{tdx} or AMD SEV-SNP~\cite{amd-sev}).

TEEs serve two purposes in \atlas.
First, TEE hardware-enforced memory integrity detects runtime tampering with
the attestation client and ML system components by the MLaaS host
(including privileged software like the OS or hypervisor).
Second, TEEs provide a hardware-based root of trust for provenance and pipeline
metadata.

\subsubsection{Artifact Measurements}\label{sec:framework:transparency:artifact}

For every artifact that is ingested into and output by an ML pipeline, the
attestation client computes a cryptographic measurement using a collision
resistant hash function resulting in a unique, immutable identifier.
If an artifact is tampered with, its measurement will differ from its golden value, allowing \atlas verification services to detect modifications between lifecycle stages (\textbf{R1}).
Artifact producers are expected to publish digitally signed measurements as
golden values whenever an artifact is first created.

\subsubsection{Model Transformation Integrity}\label{sec:framework:transparency:transformation}

The attestation client is also responsible for generating provenance and
pipeline metadata describing the transformation process and ML system that
produced a new model artifact.
Before pipeline execution begins, the client obtains a hardware attestation to its
initial state from its TEE, which includes measurements
of the client's execution environment.

At the start of pipeline execution, the attestation client remotely attests the
ML system's TEE~\cite{tdx,amd-sev}, verifying the integrity of the \emph{pipeline's}
compute environment.
Specifically, the client checks that the ML system's
firmware, OS and pipeline code match the golden values published by the MLaaS
provider to ensure that a pipeline starts from a known good state.\footnote{Because only select TEE implementations~\cite{tdx-connect}
	provide built-in support for attested interactions with I/O devices or
	ML accelerators like GPUs, it is challenging to distinguish between benign
	and malicious \emph{runtime} modifications to
	pipelines via the network, disk, etc (see~\S\ref{sec:discussion}).}

Throughout pipeline execution, TEE hardware enforces memory integrity checks
and isolation of executing attestation client/ML system code and in-memory data,
reducing the risks of interference by compromised MLaaS providers or any co-located
ML systems (\textbf{R4}).
Further, the attestation client continuously monitors an ML system's execution,
which allows it to determine when artifacts move into or outside of the system, 
and to collect information about operations that transform the input artifacts
(see~\S\ref{sec:implementation}).
For example, during data processing, the client tracks dataset modifications and
preprocessing operations; during model training, \atlas captures state changes
in model weights, hyperparameters, and configurations.

When pipeline execution concludes, the attestation client generates
pipeline metadata containing all collected ML system runtime information
and the ML system's TEE attestation.
Then, the client creates provenance metadata including
\begin{enumerate*}[label=\arabic*)]
	\item artifact measurements,
	\item operations producing the outputs,
	\item TEE attestation for the client,
	\item pipeline metadata.
\end{enumerate*}

The attestation client then digitally signs the provenance metadata with keys it
generates within the TEE, cryptographically binding ML artifacts to the pipeline and precursor 
artifacts that created them in a \emph{transformation attestation} (\textbf{R2}).
The client uploads this attestation to the transparency service.

\subsubsection{Provenance Chains}\label{sec:framework:transparency:provenance}

To enable ML model provenance tracking throughout all of its
lifecycle stages, the attestation client
embeds the cryptographic hash of precursor artifact attestations into every
artifact's transformation attestation.
These hash values are digitally signed as part of the transformation attestation,
enabling detection of unexpected/malicious modifications between attested
ML artifact transformations.
Thus, \atlas attestation clients establish an authenticated, verifiable
\emph{provenance chain} representing a model's lineage relationships
(\textbf{R3}).

\subsection{Transparency Log}\label{sec:framework:transparency:log}

The transparency service's log makes all published golden values and
attestation client-generated metadata available
to verification services.

To enable efficient insertion and provenance verification while
accommodating the cyclical nature of the ML lifecycle, \atlas relies on two
data structures (\textbf{R6}).

First, to provide cryptographic tamper-evidence for the stored values, the
transparency log is constructed using an \emph{append-only} Merkle tree~\cite{merkle-tree1987},
meaning that pipeline metadata can be efficiently inserted in the right-most
empty leaf node of the tree (e.g., as in~\cite{sigstore2025}).
Second, to enable more efficient verification of provenance \emph{across}
pipelines (or even cycles of the ML lifecycle), \atlas can represent each
discrete pipeline/cycle using a different Merkle tree. 
These separate trees are linked by embedding the Merkle root hash of the preceding
pipeline or cycle into the ``latest'' Merkle tree structure (e.g., as
in~\cite{coniks2015,hu2021merkle2}), providing a temporal cryptographic
\emph{tree chain}.
Optimizations to chained Merkle trees have been developed in prior research~\cite{hu2021merkle2,malvai2023parakeet,sun2024ats}.

\subsection{Verification Service}\label{sec:framework:verification}

Stakeholders in \atlas seek to validate that artifacts
have not been tampered with and that they were produced by expected pipelines
running in high-integrity execution environments.
To respond to verification requests, the \atlas verification service obtains golden values
and transformation attestations from transparency logs relevant to an artifact in question.

First, the verification service validates the digital signatures on
attestations and golden values to authenticate their producers. It then checks whether the artifact matches its golden value.
If these checks pass, the service inspects the
transformation attestations to confirm the ML system and
pipeline operations ran as expected based on TEE attestations
and golden values.
\atlas validates artifact lineage by traversing the provenance chain, enabling efficient verification through batching related artifact
types and maintaining a cache of verified transformations. This avoids repeated inspection of unchanged artifacts (\textbf{R6}),
which particularly benefits iterative ML pipelines.

\section{Implementation}\label{sec:implementation}

Our proof-of-concept implementation integrates with PyTorch~\cite{pytorch} and Kubeflow~\cite{kubeflow-pipelines}
through standard APIs for metadata tracking and execution monitoring within ML
pipelines.
This integration approach enabled us to avoid significant modifications to our
case study pipelines (\S\ref{sec:eval}), while maintaining \atlas' security
and transparency enhancements.
We leverage Intel TDX~\cite{tdx}, a virtualization-layer TEE, to provide the hardware-based security primitives for
ML systems and attestation clients in \atlas.

The attestation client is implemented as two components. First, a continuous ML system monitor integrates with PyTorch to collect metadata for a given pipeline.
Second, the metadata sidecar (\S\ref{sec:implementation:metadata}) running inside
a dedicated Intel TDX TEE generates ML artifact and
pipeline metadata in C2PA manifest format~\cite{c2pa2024spec}.

Due to current poor support for automated C2PA manifest generation for ML models, we implemented a Rust-based library and CLI\footnote{Available at github.com/IntelLabs/atlas-cli}
that captures artifact measurements, Intel TDX attestations, and digital
signatures in C2PA format.
Supporting other software provenance formats~\cite{slsa2025, spdx-build2023}
is future work.

We extend Sigstore's Rekor~\cite{sigstore2025} to support \atlas{} C2PA-based model transformation attestations,
validating signatures and measurements to ensure only properly signed artifacts are stored.

Out of space considerations, we provide additional details about the implementation
in App.~\ref{sec:appendix:impl}.

\Paragraph{\atlas Workflow Example.}
 We illustrate \atlas's end-to-end operations through an example with fine-tuning a BERT model for sentiment analysis:

 \begin{enumerate}[nosep]
    \item \textbf{Pipeline Environment Provisioning:} MLaaS provider sets up \atlas{} attestation client and ML system monitor in Kubeflow.
    \item \textbf{Data Preparation:} Data scientist prepares and uploads custom dataset, with \atlas{} metadata sidecar measuring and attesting the dataset using Intel TDX, submitting attestation to the transparency log.
    \item \textbf{Environment Validation:} \atlas{} sidecar verifies training environment integrity, adding TDX-based attestation to the C2PA manifest.
    \item \textbf{Training Process Monitoring:} Attestation client tracks (App.~\ref{sec:appendix:monitoring}):
    \begin{itemize}
    \item Model weight changes via PyTorch hooks
    \item Checkpoint creation and modifications
    \item Hyperparameter updates
    \end{itemize}
    \item \textbf{Model Verification:} Third-party regulator verifies model provenance using \atlas{} verification service.
    \item \textbf{Deployment:} Model vendor deploys verified model with provenance chain for user and application integrity validation.
    \end{enumerate}

\noindent\textbf{Metadata Sidecar.}\label{sec:implementation:metadata}
Because the \atlas ML system monitor runs alongside untrusted MLaaS provider code,
the attestation client's metadata sidecar leverages TEE remote attestation to
detect tampering with the ML system monitor.
That is, the sidecar interfaces with the ML system to obtain the Intel TDX-based
compute environment attestations that capture TEE state and ML system component
measurements, which are cryptographically anchored in hardware.
We use the Confidential Containers (CoCo) framework~\cite{coco2024attestation}
to implement the remote attestation procedure in the sidecar and ML system monitor.

Once the ML system monitor's integrity has been validated, the sidecar generates and
digitally signs C2PA manifests.
These include the sidecar's and ML system's Intel TDX attestations, the received ML
system metadata, the measurements for the pipeline's input and output artifacts,
and hashes for any linked transformation attestations.
We describe a storage optimization in App.~\ref{sec:appendix:impl:cache}.

\noindent\textbf{Verification Service Implementation.}\label{sec:implementation:verification}
For ease of implementation, the metadata sidecar also serves as a verification endpoint allowing pipeline components to validate artifact
integrity against stored attestations.
We optimize the performance of our staged verification system in three ways:
1) by processing changes incrementally and caching to avoid re-verifying
unchanged components, 2) via batch processing of verification operations, and
3) parallel verification paths for independent component classes.
App.~\ref{sec:appendix:impl:verif-optimizations} provides additional details.

\section{Evaluation}\label{sec:eval}

We validate our framework through a security analysis, preliminary performance testing 
and a case study with a BERT Meta~\cite{lin2023metabert, lin2023metabertimpl} fine-tuning pipeline.

\subsection{Security Analysis}\label{sec:eval:security}
\atlas provides measures against the threats outlined in~\S\ref{sec:problem:threat-model} through multiple security mechanisms.

For MLaaS provider threats, the hardware-rooted TEEs in \atlas isolate sensitive computations and
detect malicious insider tampering with executing ML pipelines.
The attestation client continuously validates the runtime environment,
generating ML system measurements that are cryptographically bound to model artifacts.

\atlas{} counters hub threats by verifying artifact integrity through cryptographic measurements and signatures, maintaining a provenance chain
that identifies mismatched signatures or tampered dependencies before they propagate.

\atlas{} mitigates artifact producer threats through comprehensive provenance tracking,
providing an immutable record of pipeline operations that detects undeclared dependencies
and intentional omissions.

\subsection{Preliminary Performance Analysis}\label{sec:eval:performance}

We conduct our experiments on Intel\textsuperscript{\textregistered} Xeon\textsuperscript{\textregistered} Gold 5520+ processors with 256 GB
of RAM running Ubuntu 24.04 beta. Employing \atlas{} with the BERT Meta case study's CPU-only PyTorch-based fine-tuning pipeline, where
the provenance chain covers $20$ artifacts (up to $120$ for more complex pipelines).
Our measurements demonstrate near-linear scalability of verification time across different chain lengths and model sizes.

Preliminary tests show \atlas{} security mechanisms introduce minimal
training overhead (under 8\%), with each C2PA provenance manifest (8KB) containing artifact measurements, TEE attestations, and pipeline metadata.
Verification processes scale linearly with model size, and our caching strategies reduce verification latency by up to 50\%, achieving near-constant
time for cached (see~\S\ref{sec:implementation}) component verification.

For large-scale operations, performance is maintained through concurrent verification operations,
cache optimization, and selective invalidation for error handling.
We plan to conduct more extensive benchmarking in future work.

\subsection{Case Study}\label{sec:eval:casestudies}
BERT was selected for its complex architecture and widespread production use. Our implementation
covers the complete lifecycle from pre-trained model fine-tuning through deployment, with Intel TDX TEEs for attention computations and weight updates. 
\atlas{} secures instruction-based configuration using JSON records with query-positive-negative text triplets, tracking model adaptations
and maintaining verifiable records of hyperparameters and training progressions.

The BERT Meta implementation demonstrates performance consistent with our analysis in~\S\ref{sec:eval:performance}.

\subsection{Discussion \& Limitations}\label{sec:discussion}
Our implementation reveals limitations in current hardware security. While \atlas provides TEE-based protection for CPU operations~\cite{tdx},
ML workloads rely on GPUs and accelerators without equivalent security features~\cite{intel2024tdx}, creating security and trust boundaries between protected
and unprotected environments~\cite{menetrey2022attestation}. Emerging solutions like confidential GPU computing show promise but have performance
trade-offs~\cite{azure-h100-preview, h100-benchmark}. 

Additionally, ML lifecycle transparency creates competing requirements between verification and confidentiality of intellectual property and sensitive data~\cite{tee-survey}, an important challenge that \atlas needs to address in a future version.
Organizations must balance security requirements with operational efficiency, considering factors like verification
frequency, attestation depth, and computational overhead.

\section{Conclusion}\label{sec:conclusion}

The combination of hardware-backed security with runtime provenance tracking in \atlas provides a foundation for securing ML pipelines.
Our case study shows \atlas's ability to integrate into existing ML frameworks with reasonable performance.
Interesting directions for future work include:
\begin{enumerate*}
    \item provenance tracking of ML accelerator-based computations,
    \item end-to-end ML lifecycle confidentiality, and
    \item algorithmic verification methods and model guardrails against attacks targeting model behavior.
\end{enumerate*}

\bibliographystyle{IEEEtran}
\bibliography{references}

\begin{appendices}\label{sec:appendix}
	
	\section{Background \& Related Work}\label{sec:appendix:related}
	In addition to the works highlighted in~\S\ref{sec:background-related}, we
	describe further details and approaches about related work addressing ML
	lifecycle security and integrity.
	
	\subsection{Data Provenance \& Authenticity}
	
	\Paragraph{C2PA.} The Coalition for Content Provenance and Authenticity (C2PA)
	specification~\cite{iso2024c2pa, c2pa2024spec} was initially introduced by
	the Content Authenticity Initiative (CAI)~\cite{adobe2019cai} and Project
	Origin~\cite{project-origin} as a response to the growing challenge of
	deepfakes~\cite{adobe2019cai} and digital content manipulation,
	gaining traction in digital photography and journalism workflows.
	
	\Paragraph{LakeFS.} LakeFS~\cite{lakefs2025} combines Git-like semantics with
	concepts from object stores such as S3 to provide a version control system
	for data, including ML datsets.
	Thus, LakeFS aims to capture data lineage by tracking changes to stored data
	over time, and allowing ML applications to reference specific versions of the
	stored data.
	This approach is meant to integrate with existing first-party data processing
	pipelines, but does not facilitate verification of data provenance by downstream
	consumers.
	\atlas' metadata centered approach, on the other hand, enables first- \emph{and}
	third-party ML dataset consumers to track changes and check their provenance,
	even when they may not have direct access to the data.
	
	\subsection{Supply Chain Integrity}

	\Paragraph{BOM.}
	Bills of Materials (BOM) have been employed to document the list of components
	of a hardware or software product for over three decades~\cite{hegge191bom}.
    Software BOM (SBOM) have been the focus of many industry and academic efforts seeking to
	facilitate tracking software dependencies and other metadata~\cite{ntia-sbom},
	to improve their adoption, and to enhance SBOM integrity and privacy (e.g.,~\cite{zahan2023,ntia-suppliers2021}).
	
	Similarly, the AIBOM framework~\cite{trail-of-bits-aibom2024, manifest-aibom2023}
	focuses on intended ML model supply chain management.
	Like SBOM, AIBOM provide a mechanism for tracking model software dependencies
	and maintaining model metadata.
	At the time of writing, we are not aware of any frameworks other than \atlas
	that utilize any sort of BOM data format to track ML model components.
	
	\Paragraph{Authenticated claims.}
	A number of frameworks for capturing and verifying a variety of security
	claims and metadata about the supply chain have been proposed.
	in-toto~\cite{torres2019} collects authenticated claims \emph{across} supply
	chain steps, including SBOM and SLSA metadata.
	In particular, in-toto enables software development pipeline owners and
	downstream artifact consumers to specify end-to-end supply chain policies,
	and validate that only the expected parties carried out specific steps in the
	pipeline and artifacts underwent transformations in the expected order.
	Given recent and upcoming enhancements that further
	generalize the framework, in-toto may be a suitable option for specifying and
	verifying end-to-end ML model pipeline integrity policies in \atlas.
	
	Sigstore~\cite{sigstore2025} provides a transparency log-based infrastructure
	for issuing signing credentials and validating digital signatures on supply
	chain artifacts and metadata.
	Similarly, Supply Chain Integrity, Transparency and Trust
	(SCITT)~\cite{scitt2024} is an architecture for implementing distributed ledger
	based supply chain integrity mechanisms, providing global visibility and auditing
	for supply chain operations and claims.
	The SCITT architecture also includes confidential computing technologies that
	help ensure that only authorized parties submit claims to the transparency
	ledger.

    \section{Implementation Details}\label{sec:appendix:impl}

    \subsection{Kubeflow Integration}\label{sec:appendix:impl:pipeline}
    The integration with Kubeflow is achieved through custom operators and controllers that monitor pipeline
    execution through Kubeflow's Metadata V2 Beta API and KFP API. Through the \texttt{//apis/v2beta1/metadata} endpoint,
    we track execution contexts and maintain verifiable records of pipeline runs.

    By interfacing with \texttt{/apis/v2beta1/artifacts}, we track model artifacts and their lineage. The
    metadata store provides structured information about component dependencies and data flow through the
    \texttt{/apis/v2beta1/connections} endpoint. Our system correlates this information with integrity measurements and hardware
    attestations, creating verifiable records of pipeline execution states.

    The metadata extraction leverages Kubeflow's event system through \texttt{/apis/v2beta1/events}, enabling
    real-time capture of pipeline state transitions, component execution details, artifact generation events, and parameter
    updates. This structured approach enables verification of pipeline states while maintaining compatibility with
    existing workflows.
        
    \subsection{C2PA Metadata Examples}\label{sec:appendix:impl:examples}
    A typical execution record captured by our system looks like:
    {\footnotesize
    \begin{verbatim}
    {
        "execution": {
            "name": "training-run-132",
            "state": "RUNNING",
            "pipeline_spec": {
                "parameters": {
                    "learning_rate": 0.001,
                    "batch_size": 32,
                    "random_seed": 42,
                    "optimizer_config": {
                        "type": "Adam",
                        "beta1": 0.9,
                        "beta2": 0.999
                    }
                },
                "runtime_config": {
                    "gcs_output_directory": "gs://...",
                    "tensorflow_version": "2.9.0"
                }
            }
        }
    }
    \end{verbatim}
    }
    
    For each execution, our system adds corresponding integrity measurements and verification records:
    {\footnotesize
    \begin{verbatim}
    {
        "integrity_measurement": {
            "component_id": "training-run-132",
            "tdx_quote": "base64:...",
            "environment_hash": "sha256:...",
            "timestamp": "2024-01-15T10:30:00Z",
            "parameter_hash": "sha256:..."
        }
    }
    \end{verbatim}
    }

    \subsection{ML System Monitoring Procedures}\label{sec:appendix:monitoring}

    Our proof-of-concept implementation leverages several techniques to monitor ML pipeline activities with minimal intrusion into existing workflows. The implementation focuses on collecting runtime data about model weights, hyperparameters, and execution context while maintaining performance and compatibility with established ML frameworks.

	\subsubsection{File System Monitoring in the Atlas Framework}\label{sec:appendix:monitoring:fs}
	
	The Atlas sidecar implements a file system monitor written in Rust that detects checkpoint creation and modification events:
	
	\begin{algorithmic}[1]
		\Procedure{InitCheckpointMonitor}{$dir, client$}
		\State Initialize directory and client references
		\State Create empty checksum tracking map
		\State Setup file system watcher
		\EndProcedure
		
		\Procedure{SetupWatcher}{}
		\State Create event listener for file changes
		\State Start background monitoring thread
		\State Register directory for change notifications
		\EndProcedure
		
		\Procedure{ScanCheckpoints}{}
		\For{each checkpoint file in directory}
		\State Compute file cryptographic checksum
		\State Store checksum in tracking map
		\State Register existing checkpoint in metadata
		\EndFor
		\EndProcedure
		
		\Procedure{OnFileCreated}{$file$}
		\If{file is checkpoint type}
		\State Compute checksum and record creation
		\State Update checksum tracking map
		\EndIf
		\EndProcedure
		
		\Procedure{OnFileModified}{$file$}
		\If{file is checkpoint type}
		\State Compute new checksum
		\State Retrieve old checksum from map
		\If{checksums differ}
		\State Update tracking map
		\State Record modification in metadata
		\EndIf
		\EndIf
		\EndProcedure
	\end{algorithmic}

    \subsubsection{Callback/Hook Registration in PyTorch}\label{sec:appendix:monitoring:hooks}

    For our BERT Meta case study, we implemented a callback system that integrates with PyTorch's event mechanisms. The model monitoring component operates as follows:

    \begin{algorithmic}[1]
    \Procedure{InitModelMonitor}{$model, client$}
      \State Store references to model and client
      \State Register monitoring hooks on model
    \EndProcedure

    \Procedure{RegisterHooks}{}
      \For{each layer module in model}
        \If{module is neural network layer}
          \State Attach forward hook for activation capture
          \If{module has trainable weights}
            \State Attach gradient hook for updates
          \EndIf
        \EndIf
      \EndFor
    \EndProcedure

    \Procedure{ForwardHook}{$module, input, output$}
      \State Calculate unique layer identifier
      \State Extract statistical metrics from output
      \State Record activation data to metadata store
    \EndProcedure

    \Procedure{GradientHook}{$gradient$}
      \State Calculate gradient magnitude
      \State Record gradient event with timestamp
    \EndProcedure
    \end{algorithmic}

    Additionally,for some cases we extended PyTorch's standard training loop with epoch-level callbacks~\cite{pytorchlightning2025callbacks}:

    \begin{algorithmic}[1]
    \Procedure{InitTrainingCallback}{$client$}
      \State Store reference to metadata client
    \EndProcedure

    \Procedure{OnEpochStart}{$epoch, optimizer$}
      \State Create hash of optimizer configuration
      \State Record epoch start event with config hash
    \EndProcedure

    \Procedure{OnEpochEnd}{$epoch, metrics, model$}
      \State Capture cryptographic model state snapshot
      \State Record completion with metrics and snapshot
    \EndProcedure
    \end{algorithmic}

    These hooks operate with minimal overhead while providing comprehensive visibility into the model's evolution.

    \subsubsection{Configuration Wrappers in the Atlas Framework}\label{sec:appendix:monitoring:config}
    
    To extract hyperparameter access and modifications from PyTorch, we implement transparent wrapper classes:

    \begin{algorithmic}[1]
    \Procedure{InitTrackedConfig}{$config, client$}
      \State Store config and client references
      \State Initialize version counter to zero
      \State Record initial configuration state
    \EndProcedure

    \Procedure{Get}{$key$}
      \State Log access event to metadata store
      \State Return value for requested key
    \EndProcedure

    \Procedure{Set}{$key, value$}
      \State Retrieve current value for key
      \State Update configuration with new value
      \State Increment version counter
      \State Record modification in metadata
      \State Update configuration state hash
    \EndProcedure

    \Procedure{RecordState}{}
      \State Generate hash of current configuration
      \State Store versioned snapshot in metadata
    \EndProcedure

    \Procedure{GetConfig}{}
      \State Return copy of configuration
    \EndProcedure
    \end{algorithmic}

    \subsection{Integration Between Framework and ML Pipeline}\label{sec:appendix:monitoring:integration}

    For the BERT Meta case study, we developed an integration layer that allows the framework components to interact with the Python ML pipeline:

    \begin{algorithmic}[1]
    \Procedure{InitBridgeService}{$endpoint, dir$}
      \State Create metadata client connection
      \State Initialize checkpoint monitor
      \State Start bridge service
    \EndProcedure

    \Procedure{InitConfig}{$configJSON$}
      \State Parse configuration from JSON
      \State Create configuration tracking wrapper
      \State Generate unique tracking identifier
      \State \Return tracking identifier
    \EndProcedure

    \Procedure{ScanCheckpoints}{}
      \State Scan and register existing checkpoints
    \EndProcedure

    \Procedure{RecordEvent}{$type, data$}
      \If{$type$ is epoch start}
        \State Record epoch initialization
      \ElsIf{$type$ is epoch end}
        \State Record epoch completion with metrics
      \ElsIf{$type$ is layer activation}
        \State Record layer output statistics
      \ElsIf{$type$ is gradient event}
        \State Record gradient flow information
      \EndIf
    \EndProcedure
    \end{algorithmic}

    On the Python side, we implement a complementary bridge client:

    \begin{algorithmic}[1]
    \Procedure{InitBridgeClient}{$socketPath$}
      \State Connect to monitoring bridge service
    \EndProcedure

    \Procedure{SetupMonitoring}{$model, config, dir$}
      \State Serialize configuration to JSON
      \State Initialize configuration tracking
      \State Scan existing model checkpoints
      \State Create model monitoring hooks
      \State Setup training loop callbacks
      \State \Return monitoring components
    \EndProcedure

    \Procedure{RecordLayerData}{$layer, stats$}
      \State Package activation statistics
      \State Send to bridge as layer event
    \EndProcedure

    \Procedure{RecordGradient}{$magnitude, time$}
      \State Package gradient information
      \State Send to bridge as gradient event
    \EndProcedure

    \Procedure{RecordEpochStart}{$epoch, optHash$}
      \State Package epoch initialization data
      \State Send to bridge as epoch start event
    \EndProcedure

    \Procedure{RecordEpochEnd}{$epoch, metrics, hash$}
      \State Package completion metrics
      \State Send to bridge as epoch end event
    \EndProcedure
    \end{algorithmic}

    \begin{figure}[h]
        \centering
        \includegraphics[width=0.5\textwidth]{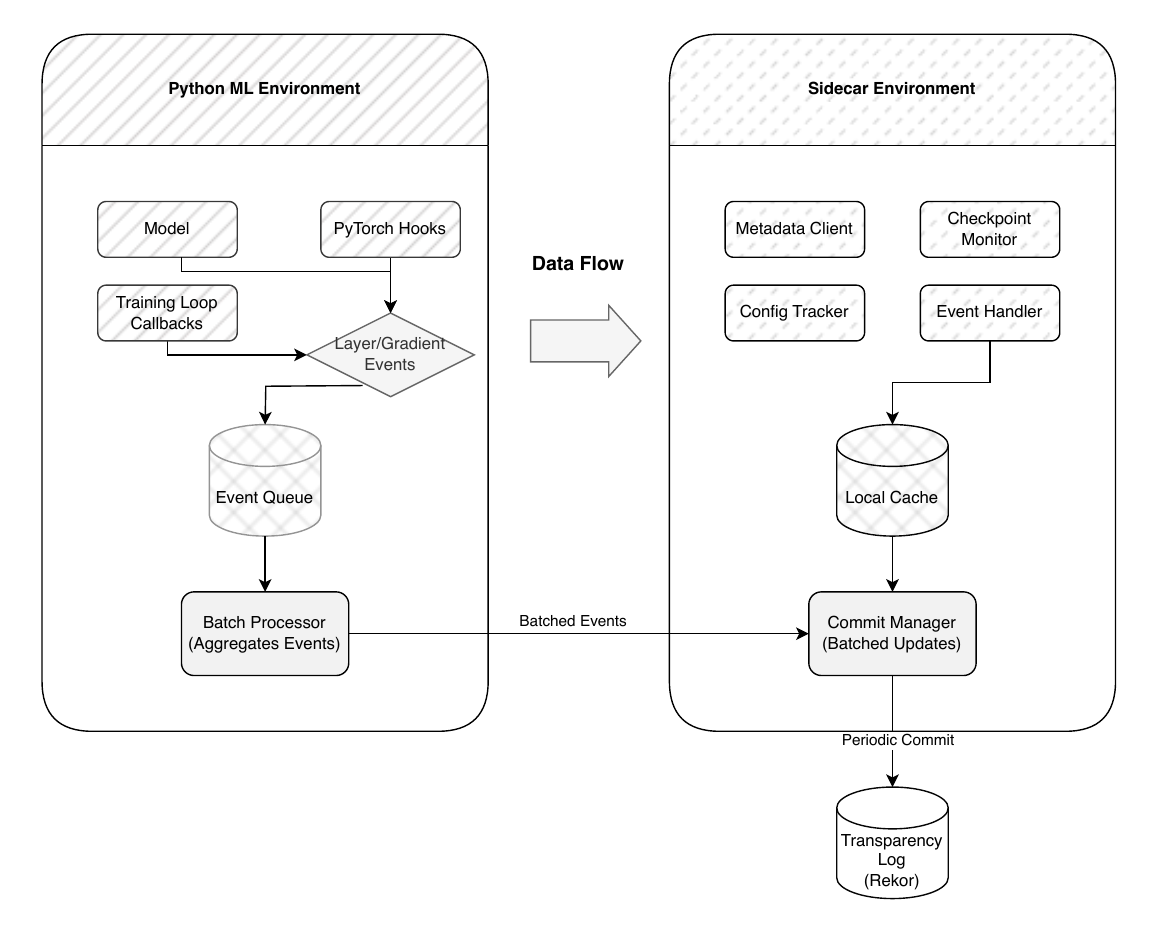}
        \caption{Altas sidecar collector showing data flow between Python ML environment and framework: The diagram illustrates how monitoring events are cached before being committed to the transparency log.}
        \label{fig:atlas-bridge-architecture}
       \end{figure}
    This hybrid architecture enables monitoring of the BERT training process with minimal modifications to the existing pipeline, while leveraging efficient system-level operations for the monitoring infrastructure. 
    
    \subsection{Attestation Client Storage Optimization}\label{sec:appendix:impl:cache}
    As a storage optimization, the attestation client's metadata sidecar first
    stores all generated C2PA manifests in a local cache layer before being committed to the transparency log.
    The local cache maintains an indexed hierarchy of manifests for efficient validation during pipeline execution,
    before final storage in Rekor for tamper-evident provenance tracking.
    
    More specifically, we decompose manifests into constituent components within the cache.
    The C2PA metadata assertions, claim signatures, and pipeline metadata are stored separately, with relationships maintained through a reference system.
    This approach enables efficient updates to specific manifest components, reduced storage redundancy,
    optimized query performance, and scalable version tracking.
    
    \subsection{Verification Service Optimizations}\label{sec:appendix:impl:verif-optimizations}
    
    \begin{figure}[h]
    	\centering
    	\includegraphics[width=0.5\textwidth]{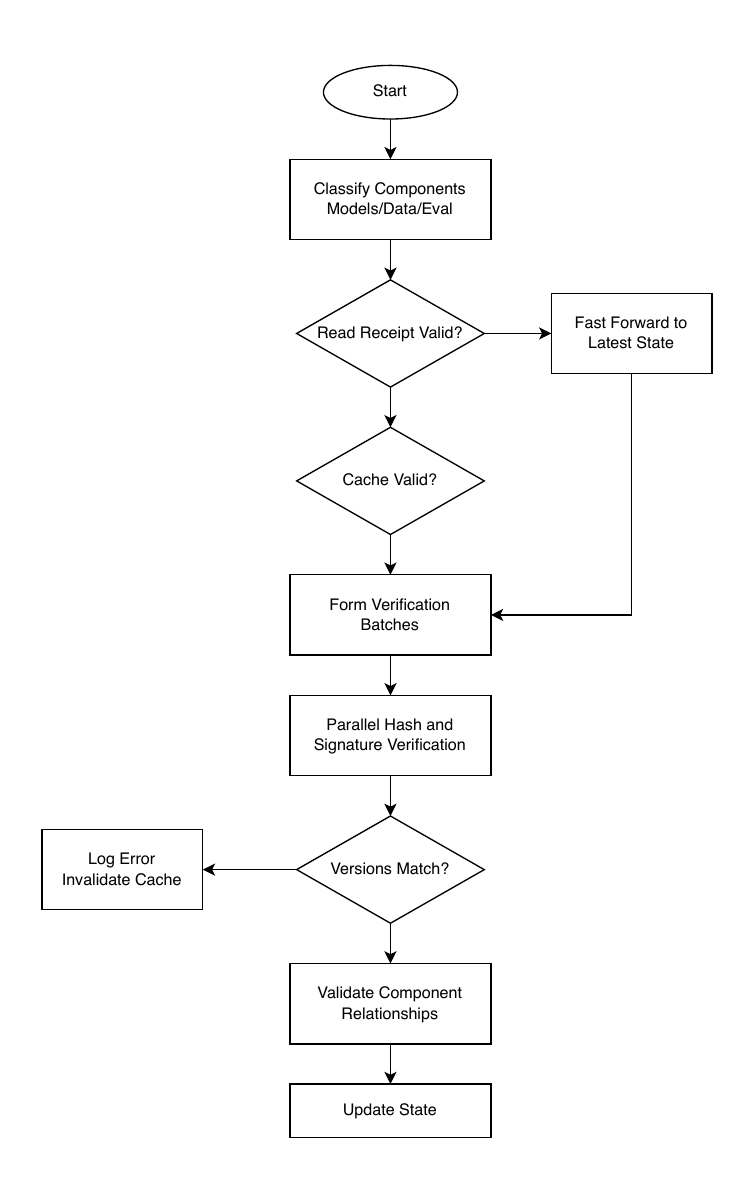}
    	\caption{Verification workflow implementation for Atlas components. The system classifies artifacts and processes changes incrementally, preserving cached states for unchanged components. Related artifacts are grouped for batch processing, with parallel validation of relationships. During errors, only affected components are invalidated, reducing verification overhead while maintaining security guarantees.}
    	\label{fig:verification-flowchart}
    \end{figure}
    
    For provenance chain validation, our verification service implementation parallelizes verification of
    cross-component dependencies, version compatibility and evaluation result consistency.
    The system also maintains verification checkpoints that serve as trusted reference points,
    enabling partial verification from the last known good state instead of complete chain recomputation.
    
    Error handling focuses on computational efficiency through targeted cache invalidation rather than complete cache clearing.
    When verification failures occur, the system preserves verified states while ensuring security through selective invalidation.
    Preliminary testing shows these optimizations reduce verification time by up to 50\% through parallelization and caching, while maintaining security guarantees.

    \subsection{Framework Adaptability}\label{sec:appendix:impl:adapt}
    The BERT deployment validated our framework's flexibility across different ML environments. The abstraction layer successfully handled variations in
    framework-specific interfaces, from PyTorch's hook mechanisms to TensorFlow's Keras callbacks, while maintaining consistent security guarantees.
    Custom adapters enabled integration without modifying existing ML infrastructure, demonstrating the framework's ability to enhance security while preserving established workflows.

    \section{Adoption Considerations}\label{sec:appendix:discussion} 
    \subsection{Storage Optimizations}\label{sec:appendix:discussion:storage}
    \subsubsection{Storage and Scalability}\label{sec:appendix:discussion:storage:scalability}
    Our analysis suggests opportunities for optimizing manifest storage through decomposed and hybrid architectures. Rather than 
    storing complete manifests as single documents, separating components like assertions, signatures, and metadata could improve 
    efficiency while maintaining security guarantees. Organizations should consider distributed storage strategies that balance 
    immutability requirements with query performance needs.

    \subsubsection{Decomposed Storage Model}\label{sec:appendix:discussion:storage:decomposed}
    Instead of storing complete manifests as single documents, the system could decompose manifests into their constituent components.
    The assertion store, claim signatures, and metadata could be stored separately, with relationships maintained through a reference system.
    This approach would enable more efficient updates and queries of specific manifest components.
    
    The decomposed model could leverage distributed ledger technology (DLT)~\cite{geeksforgeeks2023dlt} for critical manifest components
    while maintaining bulk data in optimized storage systems.
    This hybrid approach would:
    \begin{itemize}
    \item Store cryptographic proofs and signatures on the distributed ledger for immutability
    \item Maintain manifest metadata and relationships in graph databases for efficient querying
    \item Use object storage for large artifacts like model weights and datasets
    \item Link components through cryptographic references preserved in the ledger
    \end{itemize}
    
    \subsubsection{Hybrid Storage Architecture}\label{sec:appendix:discussion:storage:hybrid}
    A hybrid approach could maintain critical verification data in Rekor for its transparency guarantees
    while storing detailed manifest data in optimized storage systems.
    This would balance the need for immutable proof of existence with efficient data access and management.
    
    These storage optimizations could significantly reduce operational overhead while maintaining the security guarantees of our framework.
    Performance testing indicates potential reduction in storage requirements and query latency through these alternative approaches.
    
    \subsection{Deployment Guidelines}\label{sec:appendix:deployment}
    \subsubsection{Organizational Adaptations}\label{sec:appendix:deployment:org}
    Organization-specific adaptations are necessary to align with existing infrastructure and security policies.
    Key considerations include:
    \begin{itemize}
       \item Integration with current MLOps platforms
       \item Alignment with existing security monitoring systems
       \item Customization of verification policies
       \item Adaptation to specific hardware security capabilities
       \item Compliance with organizational security standards
    \end{itemize}
    
    \subsubsection{Security Requirement Balance}\label{sec:appendix:deployment:security}
    Security requirement balance directly impacts operational efficiency.
    Organizations must determine appropriate verification frequencies and depth based on their risk profile and performance requirements.
    For instance, continuous hardware attestation of all pipeline components provides maximum security but introduces significant overhead.
    A more balanced approach might implement full verification at critical pipeline stages while using lightweight checks during intermediate steps.
    
    \subsubsection{Computational Overhead Management}\label{sec:appendix:deployment:overhead}
    Computational overhead management becomes crucial when scaling the framework across large ML operations.
    Our implementation shows that intelligent caching of verification results and batch processing of integrity checks can significantly reduce overhead.
    Organizations should consider:
    \begin{itemize}
    	\item Strategic placement of verification checkpoints - Organizations can tailor verification intensity based on their specific security needs and operational context. While financial or healthcare institutions might require comprehensive verification throughout their ML pipeline, research or development environments might focus verification efforts primarily on model publication or deployment stages. This flexible approach enables efficient resource utilization while maintaining appropriate security levels for each use case.
    	\item Optimization of hardware attestation frequency - By analyzing pipeline characteristics and risk patterns, attestation frequency can be tuned to concentrate on high-risk operations while reducing overhead during stable processing phases.
    	\item Efficient manifest storage and retrieval mechanisms - The system maintains an indexed store of manifests with hierarchical organization, enabling quick validation of model lineage while managing storage overhead for long-term provenance tracking. 
    	\item Parallel verification processing where possible - This approach utilizes available computational resources effectively by running verification operations concurrently when component dependencies allow.
    \end{itemize}
    
    \section{Future Work}\label{sec:appendix:future}
    Several potential enhancements could extend our framework's capabilities and applicability:
    
    \Paragraph{Distributed Training Support.}\label{sec:appendix:future:distributed}
    The current framework could be enhanced to handle multiple TEEs coordinating across training nodes, with cross-node attestation
    and verification protocols. This would require developing protocols for maintaining integrity across distributed components
    while managing the additional complexity of verifying inter-node communications and state synchronization~\cite{menetrey2022attestation}.
    
    \Paragraph{Federated Learning Compatibility.}\label{sec:appendix:future:federated}
    Our current framework could be extended to support federated learning environments,
    particularly through integration with Intel's OpenFL (Open Federated Learning)~\cite{wei2020federated} framework.
    OpenFL's architecture, which separates aggregator and participant nodes while maintaining model security,
    presents unique opportunities and challenges for provenance tracking. The framework would need to extend its attestation and
    verification protocols to handle distributed model updates while preserving the privacy guarantees inherent in federated learning.
    
    Key considerations include tracking model aggregation operations, verifying participant contributions,
    and maintaining cryptographic proofs across federation rounds. OpenFL's existing security features, including
    its support for secure aggregation and TEE integration, provide natural integration points for our provenance framework.
    The challenge lies in extending our verification protocols to handle the partial model updates and differential privacy
    mechanisms~\cite{foley2022openfl} common in federated learning scenarios.
    
    \Paragraph{Enhanced Scalability Features.}\label{sec:appendix:future:scalability}
    Support for more complex ML architectures, particularly for multi-model systems and ensemble methods,
    would expand the framework's utility. This would require developing verification protocols for model composition and
    interaction, tracking dependencies between component models, and maintaining provenance across model combinations\cite{zheng2020decent}.
    
    The framework could also be extended to support dynamic trust models, allowing for flexible trust relationships between
    different components and participants in the ML pipeline.
    
    \Paragraph{Algorithmic Security Enhancements.}\label{sec:appendix:future:algorithmic}
    Our framework's modular design allows for integration of additional algorithmic security methods to enhance pipeline protection.
    Model watermarking techniques~\cite{regazzoni2021survey, boenisch2021watermarking} could be incorporated to embed verifiable ownership proofs directly into model weights,
    providing an additional layer of provenance verification.
    These watermarks would be included in the manifest chain, creating cryptographically verifiable links between model
    versions and their origins.
    
    \Paragraph{Verification Protocol Extensions.}\label{sec:appendix:future:verification}
    Our staged verification system could be extended to support dynamic trust models.
    This would allow more flexible verification policies based on component criticality and risk levels, while maintaining our core
    security guarantees. The current implementation's classification system provides a foundation for such policy-based verification.
    
    Neural fingerprinting~\cite{cao2023fingerprinting, regazzoni2021survey} methods could extend our verification capabilities by enabling detection of unauthorized model modifications or derivatives.
    By maintaining fingerprint signatures in our provenance records, the framework could track model lineage even when traditional hash-based verification is insufficient.
    This is particularly valuable for scenarios involving fine-tuning or transfer learning.
    
    Property attestation mechanisms could verify specific algorithmic characteristics of models throughout the pipeline. For example,
    robustness guarantees~\cite{benbraiek2024robustness}, fairness metrics~\cite{liang2023fairness}, or backdoor resistance~\cite{bagdasaryan2023mithridates} could be measured and included in the manifest chain.
    These properties would enhance the framework's ability to detect subtle manipulations that might not affect model
    hashes but could impact model behavior.

    Throughout this process, \atlas provides tamper-evident records through its transparency log and TEE-based attestations.
    The verification service requires access to both the model artifacts and the cryptographic measurements in the transparency log to confirm the integrity of the complete ML pipeline,
    ensuring that no unauthorized modifications occurred during the model's lifecycle.

\end{appendices}

\end{document}